\documentclass[nofootinbib,floatfix,onecolumn,prd]{revtex4-1}

\usepackage{amsmath,amssymb,amsthm}

\usepackage{graphicx}  
\usepackage{float}     

\newcommand{\ket}[1]{\lvert #1 \rangle}

\graphicspath{{./figures/}}

\begin{document}
\title{Shortcuts to adiabaticity, unexciting backgrounds, and reflectionless potentials}

\author{Fernando C. Lombardo  $^{1}$ and Francisco D. Mazzitelli $^{2}$}


\affiliation{$^1$ Departamento de F\'\i sica {\it Juan Jos\'e
 Giambiagi}, FCEyN UBA and IFIBA CONICET-UBA, Facultad de Ciencias Exactas y Naturales,
 Ciudad Universitaria, Pabell\' on I, 1428 Buenos Aires, Argentina }
\affiliation{$^2$ Centro At\'omico Bariloche and Instituto Balseiro,
Comisi\'on Nacional de Energ\'\i a At\'omica, 
R8402AGP Bariloche, Argentina}



\begin{abstract} We analyze shortcuts to adiabaticity (STA) and their completions for the quantum harmonic oscillator  (QHO) with time-dependent frequency, as well as for quantum field theory (QFT) in non-stationary backgrounds. We exploit the analogy with one-dimensional quantum mechanics, and  the well known correspondence between Bogoliubov coefficients in the QHO and transmission/reflection amplitudes in scattering theory. Within this framework, STA protocols for the QHO are equivalent to transmission resonances, while STA in QFT with homogeneous backgrounds correspond to reflectionless potentials. Moreover, using the connection between particle creation and squeezed states, we show how STA completions can be understood in terms of the anti-squeezing operator.
\end{abstract}
\maketitle
\section{Introduction}\label{sec1}

The quantum harmonic oscillator (QHO) with time-dependent frequency~\cite{Husimi} is not only a fundamental theoretical model but also a versatile tool that continues to inspire modern research. It captures essential features of quantum control protocols, underlies the design of shortcuts to adiabaticity (STA) in atomic and optical systems~\cite{STA_review}, and~provides a framework to study particle creation processes in curved spacetime and non-stationary backgrounds~\cite{Birrell}. 
In particular, systems of coupled QHO with time-dependent frequencies play a crucial role in the description of the dynamical Casimir effect (DCE)~\cite{Dodonov_rev1,Dodonov_rev2,Dodonov_rev3,Dalvit_rev,Nation}. 
 Simplicity and universality of QHO 
make it a bridge between elementary quantum mechanics and practical implementations in 
contemporary quantum~technologies.

With this paper, 
  we honor
 Victor 
 Dodonov, whose numerous contributions to quantum optics and quantum electrodynamics have placed this model at the center of 
essential
developments. From~his early studies of coherent states~\cite{Dodonov_1974} to later explorations of nonclassical states
~\cite{Dodonov_2002}, and~his more recent studies  
on the dynamics of energy evolution~\cite{Dodonov_2025} and the analysis of adiabatic versus non-adiabatic regimes~\cite{Dodonov_2024}, 
Dodonov
has consistently shown how much insight can be extracted from this paradigmatic system, not to mention his influential work 
on the DCE~\cite{Dodonov_rev1,Dodonov_rev2,Dodonov_rev3, Klimov}. 

In this paper, we reconsider the time-dependent QHO, with particular emphasis on the construction and characterization of STA 
and STA completions~\cite{STAqft1,nico_entropy}
The point is
to clarify the role of these techniques in preventing particle excitations and~to establish a link between seemingly distinct areas: (i) the conditions under which non-exciting backgrounds arise in quantum field theory~\cite{nico_entropy,Vachaspati2022,Vachaspati2022bis}, and~(ii) the appearance of transmission resonances and reflectionless potentials~\cite{KayMoses} in one-dimensional quantum mechanics. These correspondences provide both a unifying perspective and practical tools to analyze dynamical quantum systems, where adiabatic control and excitation suppression are central~issues.

The paper is organized as follows.
Section~\ref{sec2} discusses
STA within the framework of the Lewis–Riesenfeld invariant theory~\cite{LR}. In~Section~\ref{sec3}, we analyze squeezed states, their relation to particle creation, and~the role of the anti-squeezing operator, establishing a connection with the 
Lewis--Riesenfeld   approach. 
Section~\ref{sec4} gives a brief
 ~review of the analogy between the time-dependent QHO and the scattering 
problem in one-dimensional quantum mechanics, connecting STA with transmission resonances. 
Section
~\ref{sec5} 
discusses STA in the context of quantum field theory (QFT) in homogeneous backgrounds, showing their  
relationship  
with  the existence of reflectionless potentials in 
non-relativistic
quantum mechanics. Section
~\ref{sec6} gives  
 concluding~remarks.


\section{STA for the Harmonic Oscillator with Time-Dependent~Frequency}\label{sec2}

In this Section, we discuss the existence of STA and STA completions for the QHO with time-dependent 
frequency. As
 shown, 
in~the Lewis--Riesenfeld approach,  the~consideration
of STA can naturally 
be formulated 
in terms of the Ermakov function. Throughout this Section, we closely follow  our previous 
paper
~\cite{nico_entropy}.

The dynamical equation for the harmonic oscillator with a time-dependent frequency $\omega$
~is given by
\begin{equation}\label{eq:qho}
\ddot{q} + \omega^2(t) q = 0\, ,
\end{equation}
where \( q(t) \) represents the position of the oscillator and the dot on top denotes the time ($t$) derivative. 
At~the operator level, the~position operator \( \hat{q} \), in~the Heisenberg picture, can be written in terms of the annihilation and creation operators \( \hat{a} \) and \( \hat{a}^\dagger \) as
\begin{equation}
\hat{q}(t) = q(t) \hat{a} + q^*(t) \hat{a}^\dagger\, ,
\end{equation}
where, in~this case,  \( q(t) \) is any complex solution of Equation \eqref{eq:qho} properly~normalized. The asterisk (*) 
and dagger ($\dagger$) denote the complex and Hermitian conjugates, respectively.    

Assuming that the frequency tends to constant values \( \omega_{\rm in} \) (initial) and \( \omega_{\rm out} \) 
(final)
as \( t 
\to 
- \infty \) and $+ \infty$, repectively
, and~that the oscillator is initially in the ground state \( |0_{\rm in}\rangle \) for \( t \to -\infty \), 
the~system is, 
in general, 
excited as \( t \to +\infty \), if~the evolution is non-adiabatic. In~this case,  
the `in` and `out` 
vacua are
different, and~the vacuum persistence probability is
strictly less than 1, 
\( |\langle 0_{\rm out}|0_{\rm in}\rangle|^2< 1 \). 

The 
 `in` and `out` states correspond to solutions of Equation \eqref{eq:qho} with the asymptotics
\begin{equation}\label{solnINOUT}
q^{\rm in,out}(t) \xrightarrow[t \to \pm \infty]{} \frac{1}{\sqrt{2 \omega^{\rm in,out}}} e^{-i \omega^{\rm in,out} t}\, .
\end{equation}
The 
Bogoliubov transformation connecting the
 `in` and `out` bases, as~well as the corresponding Fock spaces, 
reads
\begin{eqnarray}
q^{\rm out} &=& \alpha \, q^{\rm in} + \beta \, q^{\rm in\, *} \nonumber \\
\hat{a}^{\rm out} &=& \alpha^* \, \hat{a}^{\rm in} - \beta^* \, \hat{a}^{\rm in\, \dagger}\, .    
\label{eq:Bogo}
\end{eqnarray}
As 
is demonstrated in Section ~\ref{sec2.2} 
below, 
the~ `in` 
vacuum 
can be written as a squeezed state with respect to the `out` states.

A frequency protocol \( \omega(t) \) that results in an evolution where the system does not get excited is identified as an STA. It is 
worthy emphasizing 
that, although~the evolution in an STA is non-adiabatic at intermediate times, the~system eventually returns to its initial state once the effective frequency becomes constant as \( t \to +\infty \). 
That is,
the~system undergoes transient excitations during the non-adiabatic stage but relaxes to the ground state~asymptotically.

\subsection{The Ermakov~Equation} \label{sec2.1}

In the context of the Lewis–Riesenfeld approach~\cite{LR}, the~solutions to Equation \eqref{eq:qho} are conveniently expressed in terms of the Ermakov function. The~normalization of the mode functions is fixed by the Wronskian condition $\dot{q} q^* - \dot{q}^* q = i$, 
which allows parametrizing the solution as
\begin{equation}\label{solnW}
q(t) = \frac{1}{\sqrt{2 W(t)}} e^{i \int^t W(t') \text{d}t'}\, .
\end{equation}
Here \( W(t) \) is a real function that satisfies
\begin{equation}\label{eq:WKB}
\omega^2 = W^2 + \frac{1}{2}  \frac{\ddot{W}}{W} - \frac{3}{4} \left( \frac{\dot{W}}{W} \right)^2 \, ,
\end{equation}
which is equivalent to Equation \eqref{eq:qho}. 

For quite 
slowly varying frequencies \( \omega(t) \), one
can approximate \( W \approx \omega \), and then Equation \eqref{solnW} reduces to the standard 
 Wentzel--Kramers--Brillouin
(WKB)   
solution at leading order. Higher-order corrections can be obtained by solving Equation \eqref{eq:WKB} recursively, or~alternatively, one may use an inverse-engineering approach: given a function $W(t)$, Equation \eqref{solnW} is the exact solution for the harmonic oscillator with frequency $\omega(t)$ determined by Equation \eqref{eq:WKB}.

By defining the Ermakov function as\vspace{-6pt}
\begin{equation}\label{eq:fErmakov}
\rho = \frac{1}{\sqrt{W}} \, ,
\end{equation}
one finds that the function \eqref{eq:fErmakov}  
 satisfies the Ermakov equation 
\begin{equation}
\label{eq:Ermakov}
\ddot{\rho} + \omega^2(t) \rho - \frac{1}{\rho^3} = 0\, ,
\end{equation}
which gives a
compact 
formulation of the~dynamics. 

\begin{figure}[H]
\centering

\begin{minipage}[b]{0.32\textwidth}
    \centering
    \includegraphics[width=\textwidth]{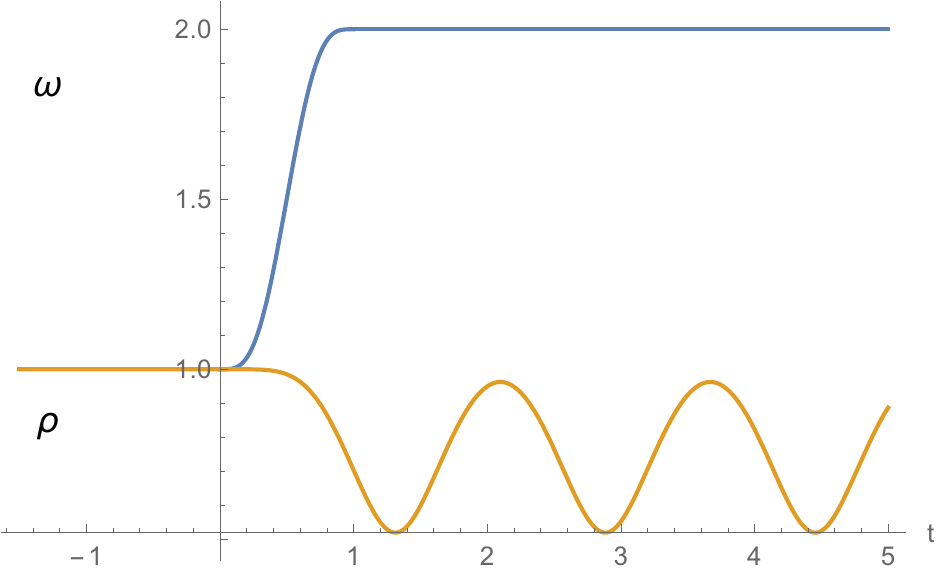}
    \\ (a)
\end{minipage}
\hfill
\begin{minipage}[b]{0.32\textwidth}
    \centering
    \includegraphics[width=\textwidth]{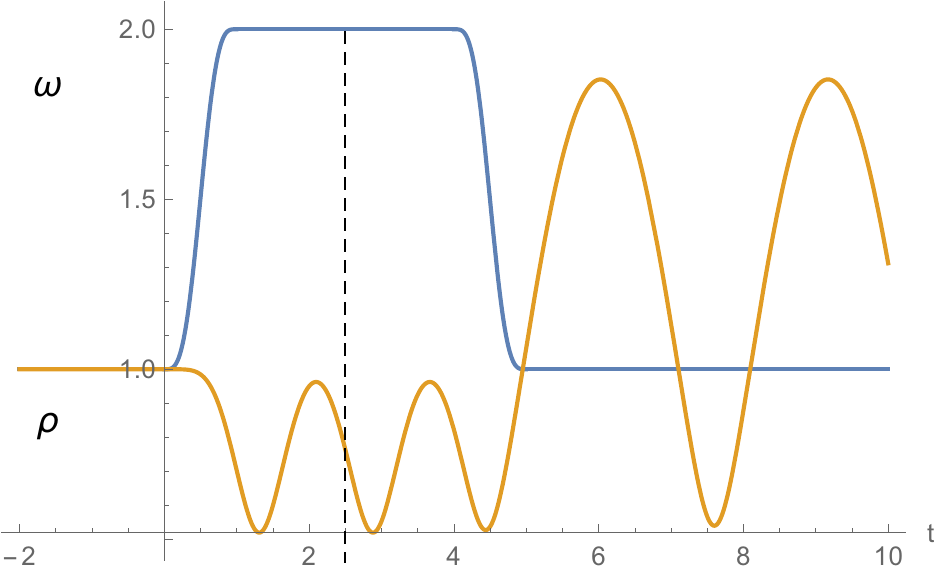}
    \\ (b)
\end{minipage}
\hfill
\begin{minipage}[b]{0.32\textwidth}
    \centering
    \includegraphics[width=\textwidth]{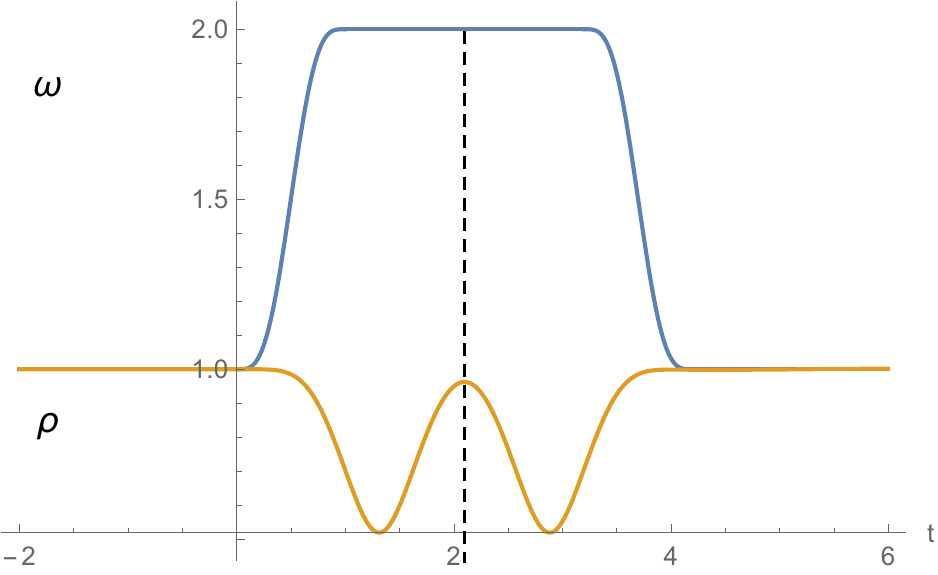}
    \\ (c)
\end{minipage}

\caption{Plot of the time dependent frequency $\omega(t)$ (blue curve) and the Ermakov function (orange curve). (a) The frequency follows a (soft) ``step-like'' evolution. The Ermakov function is initially constant and then oscillates, indicating an excited final state. (b) Similar behaviour for a "barrier-like" evolution. The dashed vertical line indicates the center of the barrier. (c) When the barrier is centered on an extremum of the Ermakov function,  the Ermakov function becomes constant, indicating a STA completion.}
\label{fig:tres}
\end{figure}

It can be shown that, for~a constant frequency \( \omega_0 \), the~general solution to the Ermakov equation 
\eqref{eq:Ermakov}
  is~\cite{Gjaja}
\begin{equation}\label{erm-asym} 
\rho^2(t) = \frac{1}{\omega_0} \left[ \cosh\delta - \sinh\delta \sin(2\omega_0 t + \varphi) \right]\, ,
\end{equation}
where \( \delta \) and \( \varphi \) are constants. For~\( \delta = 0 \), one recovers the
standard
solution of the harmonic oscillator with frequency \( \omega_0 \). Consider now the following time-dependent situation: the frequency is \( \omega_0 \) for \( t < 0 \), time-dependent in the interval \( 0 < t < \tau \), and~then takes a constant value \( \omega_1 \). 
In~this case, one has
\( \rho^2 = 1/\omega_0 \) for \( t < 0 \), and,  for~\( t > \tau \), \( \rho^2 \) 
takes
 the form of that
in Equation \eqref{erm-asym} with \( \omega_0 \to \omega_1 \), that is, an~oscillatory function.  This behavior is illustrated in Figure~\ref{fig:tres}a,b. 
 As shown in Section \ref{sec2.2}, 
this leads to a squeezed state. However, 
for time-dependent frequencies such that  \( \rho \) remains constant after the frequency becomes constant, 
one gets
an unexciting evolution,   i.e.,~ \( |\langle 0_{\rm out}|0_{\rm in}\rangle|^2=1 \). This is illustrated in Figure~\ref{fig:tres}c.

This observation allows 
 using of reverse engineering to construct effective unexciting evolutions \( \omega_{\rm eff}^2(t) \). This can be acheived by 
choosing a reference function \( \rho_{\rm ref}(t) \) that is constant both for \( t < 0 \) and \( t > \tau \),
and time-dependent in the interval $0<t<\tau$. By substituting the reference function into Equation \eqref{eq:Ermakov}, one finds
\begin{equation}
\label{eq:Ermakov2}
\omega_{\rm eff}^2 = \frac{1}{\rho_{\rm ref}^4} - \frac{\ddot{\rho}_{\rm ref}}{\rho_{\rm ref}}\, .
\end{equation}
For the
effective evolution \eqref{eq:Ermakov2},  
the~solution \( q(t) \) follows the adiabatic form 
of Equation \eqref{solnW} corresponding to the reference frequency \( \omega_{\rm ref} = 1/\rho^2_{\rm ref}\) in a finite amount of~time.

\subsection{STA Completions from the Ermakov~Function} \label{sec2.2}

Let us
now discuss whether 
the excitations generated by a given frequency evolution can be undone. 
 We refer to this process as  STA completion. Suppose that the 
oscillator has frequency \( \omega_{-} \) for \( t < 0 \), and~evolves from \( \omega_{-} \) to \( \omega_{+} \) 
within the interval
\( 0 < t < \tau_1 \), generating excitations.  
Let us 
denote this interpolating function by \( \omega_{\rm I}(t) \)  and the corresponding ground states by $ | 0_{-} \rangle $ (for $t<0$) and $ | 0_{+} \rangle $ (for $t>\tau_1$). After~this first stage, for~\( t > \tau_1 \), the~system is in a squeezed state with respect to the new basis. 
Indeed, from~the Bogoliubov transformation one can show that~\cite{Matacz,Calzetta,Lin}\vspace{-6pt}
\begin{equation}
| 0_{-} \rangle = c_0 \sum_{n \geq 0} \left( -\frac{\beta^*}{\alpha} \right)^n \frac{\sqrt{(2n)!}}{2^n n!} | 2n_{+} \rangle\, ,
\label{in-out-squeezed}
\end{equation}
where \( \alpha \) and \( \beta \) are the Bogoliubov coefficients and $n$ counts the occupation number of the `$+$` states.
The~mean occupation number of the 
`$+$` states in the initial vacuum ~is 
\begin{equation}
\langle 0_{-} | \hat a_{+}^\dagger \hat a_{+} | 0_{-} \rangle = |\beta|^2\, .
\end{equation}

In terms of the Ermakov function, the~first part of the evolution determines \( \rho(t) \) via Equation~\eqref{eq:Ermakov}. 
For~\( 0 < t < \tau_1 \), \( \rho(t) \) is fixed by \( \omega_{\rm I}(t) \), and~for \( \tau_1 < t < t_1 \) it oscillates 
according to the asymptotic form \eqref{erm-asym}. Thus, the~squeezed state generated after the first stage is 
entirelycharacterized by the corresponding Ermakov~function. 

To achieve de-excitation, one must design a second frequency evolution, \( \omega_{\rm II}(t) \), acting during the time
interval \( t_1 < t < t_1 + \tau_2 \). 
The point 
is to extend \(\rho(t)\) smoothly so that
\begin{equation}\label{eq:rcont}
\rho^2(t) = \frac{1}{\omega_{++}} 
 \quad \text{for } t > t_1 + \tau_2\, .
\end{equation}
From the
 continuation \eqref{eq:rcont}  
of \(\rho(t)\), the~corresponding frequency follows from Equation \eqref{eq:Ermakov2}. In~this way, the~complete evolution, 
given by
\begin{equation}
\omega_{\rm eff}(t) = 
\begin{cases}
\omega_- & \text{for } t < 0 , \\
\omega_{\rm I}(t) & \text{for } 0 < t < \tau_1 ,  \\
\omega_+ & \text{for } \tau_1 < t < t_1 ,\\
\omega_{\rm II}(t) & \text{for } t_1 < t < t_1 + \tau_2 ,\\
\omega_{++} & \text{for } t_1 + \tau_2 < t ,
\end{cases} \label{complete_evolution}
\end{equation}
implements an STA~completion.

A simpler scenario occurs  when \( \omega_{-} = \omega_{++} \). In~this case, one can choose the second stage as the time-reversal of the first: \( \omega_{\rm II}(t) = \omega_{\rm I}(2t_n - t) \), where \( t_n \) corresponds to a maximum or minimum of the oscillatory solution \eqref{erm-asym}. By~symmetry of the Ermakov 
equation \eqref{eq:Ermakov},  
this continuation guarantees that the complete protocol is time-symmetric around \( t_n \). The~symmetric trajectory always 
exists and restores the system to the ground state. An~example of this type
of protocol is illustrated in Figure~\ref{fig:tres}c.

Finally, let us 
remark that the time-reversal of an STA is itself an STA, as immediately follows from the Ermakov~equation 
\eqref{eq:Ermakov}.


\section{STA Completions and~Anti-Squeezing} \label{sec3}
As
discussed in Section \ref{sec2.2} above,  
in~the `out`
 basis the `in`
vacuum is represented as a squeezed state \eqref{in-out-squeezed}. The~de-excitation of the QHO then corresponds to a process in which, starting from this particular state, the~subsequent evolution may drive the system back to the vacuum~state.

For a general state, unfolding the evolution might not be possible. An~arbitrary state with the same \( |\beta|^2 \) does not necessarily lead to the vacuum state after unitary evolution. A~given value of \( |\beta|^2 \) gives only the mean occupation number but~does not provide 
 all
information about the quantum state of the~oscillator.

The results of Section \ref{sec2} can be reproduced
in terms of squeezing and anti-squeezing operators. 
Indeed, consider the vacuum state  $\vert 0\rangle$ of a quantum harmonic oscillator. We perform the following steps: first apply a squeezing operator $S(r,\theta) = \exp(r/2 (e^{-i \theta} \hat{a}^2 -  e^{i \theta} \hat{a}^{\dagger 2}))$ at time \( t = 0 \) (where $r$ denotes the squeezing parameter and $\theta$ an arbitrary squeezing angle, 
which is omitted in the following
without loss of generality). Then let the system
evolve freely under the harmonic oscillator Hamiltonian until time \( t = \tau \). Finally, apply the anti-squeezing operator \( S^\dagger(r) \) at time \( \tau \).
The final state then is

\[
\ket{\psi_{\text{final}}} = S^\dagger(r) \, U(\tau) \, S(r) \ket{0},
\]
where the free evolution operator is
\[
U(\tau) = e^{-i H \tau/ \hbar}, \quad \text{with} \quad H = \hbar \omega \left( \hat a^\dagger \hat a + \frac{1}{2} \right),
\]
where $\hbar$ is the reduced Planck constant. 
If \( S^\dagger(r) \) is applied immediately after \( S(r) \), 
the vacuum state \( \ket{0} \) is recovered:
\[
\ket{\psi_{\text{final}}} = S^\dagger(r) S(r) \ket{0} = \ket{0},
\]
due to the unitarity of \( S(r) \). 
However, if~there  is  a free evolution between \( S \) and \( S^\dagger \),  the~final state reads
\[
\ket{\psi_{\text{final}}} = S^\dagger(r) e^{-i H \tau/\hbar} S(r) \ket{0}
\]
(note that the free evolution corresponds to the third step in the frequency evolution of 
Equation~\eqref{complete_evolution}). In~this case, the~final state is not generally equal to \( \ket{0} \) 
because~$S^\dagger(r) U(\tau) S(r) \neq \mathbb{I}$, where $ \mathbb{I}$ is the identity matrix. 
This is due to the fact
that \( U(\tau) \) and \( S(r) \) do not commute. Although~\( S^\dagger(r) \) formally undoes the squeezing, it does not reverse 
the intermediate time~evolution.

This result can be understood in terms of the quadrature operators \( \hat q \) and \( \hat p \):
\[
\hat q = \frac{1}{\sqrt{2}}(\hat a + \hat a^\dagger) \quad \text{and}
\quad
\hat p = \frac{1}{\sqrt{2}i}(\hat a - \hat a^\dagger).
\]
The squeezing operator \( S(r) \) compresses the uncertainty ellipse in one direction of phase space. Then, the~free evolution operator \( U(\tau) \) performs a rotation of that ellipse by an angle \( \omega \tau \). Finally, the~anti-squeezing \( S^\dagger(r) \) attempts to apply the inverse squeezing along the original axis, but~the ellipse has been rotated, so it does not recover the original circular vacuum uncertainty. Therefore, 
one
can conclude that,  
if there is free evolution between the squeezing and anti-squeezing operations, in general
the~final state will 
not be the vacuum state. The~result depends on the time interval \( \tau \), due to the rotation in phase space induced by the 
harmonic oscillator dynamics. However, if~the evolution time \( \tau \) is chosen such that the rotation corresponds to an integer multiple of full rotations, i.e.,
$\tau_n = 2\pi n/\omega, n\in \mathbb{Z}$,
then the phase space rotation satisfies
$R(\theta) = R(2\pi n) = \mathbb{I}$.
In this case, the~squeezed state's orientation returns to its original configuration, and~the inverse squeezing operator perfectly cancels the initial squeezing, up~to a global phase factor. More explicitly, for~the vacuum state,
\begin{equation}
\ket{\psi_{\mathrm{final}}} = S^\dagger(r) U\left( \frac{2\pi n}{\omega} \right) S(r) \ket{0} \\
= S^\dagger(r) S(r) e^{-i \phi} \ket{0} = e^{-i \phi} \ket{0},
\end{equation}
where \( e^{-i \phi} \) is a global~phase.

By choosing the evolution time to be an integer multiple of the oscillator period, one can compensate the rotation in phase space caused by free evolution, allowing the inverse squeezing operator to perfectly reverse the squeezing operation and recover the initial state. These results are connected with the construction of STA completions using the Ermakov function: 
it was shown in Section \ref{sec2} 
that the time reversal symmetric STA completions can be implemented only at the particular times $t_n$ where the Ermakov 
function \eqref{eq:fErmakov}  
has minima or maxima during the intermediate free evolution.  This corresponds to the anti-squeezing operator being applied at times $\tau_n$. Non symmetric STA completions are also possible but~do not correspond to~anti-squeezing.   

The excitation of the harmonic oscillator corresponds, in~the framework of QFT, 
 to~the phenomenon of ``particle creation'', 
which
is discussed in Section~\ref{sec5} below. 

\section{The Analogous Scattering Problem in Quantum~Mechanics} \label{sec4}

The dynamics of a QHO with time-dependent frequency 
is governed, at~the classical level, by~Equation \eqref{eq:qho}.
On the other hand, the~stationary Schr\"odinger equation in one-dimensional quantum mechanics is
\begin{equation}
- \frac{d^2}{dx^2}\, \psi(x) + V(x)\, \psi(x) = E\, \psi(x),
\label{eq:Schrodinger}
\end{equation}
 with $\psi(x)$ being the wave function, $V(x)$ the potential, and~$E$ the energy (we set 
$\hbar^2/2m=1$, where $m$ is the particle mass). 

A direct correspondence between Equations \eqref{eq:qho} and \eqref{eq:Schrodinger} 
is obtained by the identifications
\begin{equation}\label{eq:mapping}
q(t) \;\longleftrightarrow\; \psi(x), 
\qquad t \;\longleftrightarrow\; x, 
\qquad \omega^2(t) \;\longleftrightarrow\; E-V(x)  .
\end{equation}
 In the  
mapping \eqref{eq:mapping}, the~evolution of the QHO with time-dependent frequency 
is mathematically equivalent to a one-dimensional scattering problem 
for a particle of energy $E$ in the potential $V(x)$. 
The analogy between the time-dependent frequency QHO and the one-dimensional 
Schr\"odinger
equation \eqref{eq:Schrodinger}
has been pointed out long time ago in Ref.~\cite{Marinov}, when analyzing pair creation induced by a time-dependent electric field. It was also described in Ref.~\cite{Audretsch} in the context of 
QFT 
in curved spacetime (for more recent 
papers, 
see~\cite{Amin, Schmidt}). 
 As is shown 
below in this Section,   
there is a direct connection between the Bogoliubov transformation \eqref{eq:Bogo}
 for the QHO and the reflection and transmission coefficients in one-dimensional quantum~mechanics.

In what follows, we 
 assume that  
$\omega_{\rm in}^2=\omega_{\rm out}^2 =\omega_0^2$
so the conditions 
$E=\omega_{0}^2$ and $V(x)\to 0$ at $x=\pm \infty$ may be chosen. 
The~general solution of Equation \eqref{eq:qho} at asymptotical times 
is given by Equation \eqref{solnINOUT}, and~the 
`in` and `out` 
solutions are connected by the Bogoliubov transformation \eqref{eq:Bogo}.
Similarly, in~the scattering problem \eqref{eq:Schrodinger}, 
the positive energy solutions satisfy, asymptotically,
\begin{equation}
\psi(x \to +\infty) = \bar t \, e^{ikx},
\end{equation}
where $\bar t$ is the transmission amplitude,
while at $x \to -\infty$ the wave function is of the form
\begin{equation}
\psi(x \to -\infty) = e^{ikx} + \bar r \, e^{-ikx},
\end{equation}
with $\bar r$ the reflection amplitude
 and $k=\sqrt E$. The~corresponding probabilities satisfy 
\begin{equation}\label{eq:amplprob}
|\bar r|^2 + |\bar t|^2 = R + T = 1 .
\end{equation}

The analogy implies a direct relation between Bogoliubov coefficients in Equation \eqref{eq:Bogo}
and scattering 
probabilities in Equation \eqref{eq:amplprob}~\cite{Audretsch, Dodonov3,Amin}: 
\begin{equation}
|\beta|^2 \;\longleftrightarrow\; \frac{R}{T}, 
\qquad |\alpha|^2 \;\longleftrightarrow\; \frac{1}{T}.
\end{equation}
That is, the presence of particle creation in the QHO corresponds to a non-vanishing reflection coefficient in the scattering problem,
while the existence of an STA (absence of excitations, $\beta=0$) 
is equivalent to a transmission resonance (i.e., $T=1, R=0$) in one-dimensional quantum~mechanics.

The analog to STA completions for the QHO is, in~the scattering theory,  the~construction of  a potential such
that, for~a given energy, $R=0$ and $T=1$. The~simplest example is that of a square potential well: the first step is a jump from $V=0$ to $V=-V_0$ and the second step the reversed jump from  $V=-V_0$ to $V=0$. If~the first step is at $x=0$, by~choosing a particular value of the position of the second step 
(for example, $x=a$), 
$R=0$ and $T=1$ may be forced. 
 This is a 
known textbook result: there are resonances in the transmission coefficient, and one has  complete
transmission $T=1$ when $a \sqrt{E+V_0} = n\pi/2$, where $n$ is a positive integer.

The ``symmetric'' construction with the Ermakov function \eqref{eq:fErmakov} 
generates this type of effect: $|\beta|^2=0$ for an effective trajectory that has a particular temporal duration. Therefore, for
~the analogous scattering problem, one  should 
 be able to prove that, due to interference effects,  $R=0$ for symmetric potentials that have a particular length. 
We can prove that this is indeed the case, although~the condition on the length $a$ 
is not as straightforward
 as just above,  
since the condition  
 depends now on the form of the~potential.

Assume the potential is 
\[ 
V(x)= \left\{
\begin{array}{ll}
      U(x), & -\infty < x \leq \-a \, \, \textrm{(region I)}; \\
     -V_0, & -a\leq x\leq a\, \, \,  \,\textrm{(region II)};\\
      U(-x),  & \, a\leq x < \infty \, \, \, \, \, \textrm{(region III)}.\
\end{array} 
\right. 
\]
Here, $U(x)\leq 0$, and 
a positive energy $E$ is assumed. The~solution of the Schr\"odinger equation \eqref{eq:Schrodinger} in region I 
is 
\begin{equation}\label{eq:SchrodingerI}
\psi_{\rm I}(x)=A \varphi(x)\, ,
\end{equation}
where $\varphi(x)$ is the solution of the  Schr\"odinger equation \eqref{eq:Schrodinger} such that 
$\varphi(x)\to \exp(i k x)$ as $x\to -\infty$. 
In~region II,  
\begin{equation}
\psi_{\rm II}(x) = C e^{i \gamma x} + D e^{-i \gamma x}\, ,
\end{equation}
with $\gamma^2=E+V_0$. In~region III, 
\begin{equation}
\psi_{\rm III}(x) = F \varphi^*(-x)\, .
\end{equation}

The conditions at $x=-a$ and $x=a$
read
\begin{eqnarray}
A \varphi(-a) &=& C e^{-i \gamma a} + D e^{i \gamma a},\nonumber\\
A\varphi'(-a) &=& i\gamma(C e^{-i \gamma a} - D e^{i \gamma a}),\nonumber\\
F\varphi^*(-a) &=& C e^{i \gamma a} + D e^{-i \gamma a},\nonumber\\
F\varphi^{*'}(-a) &=& -i\gamma(C e^{i \gamma a} - D e^{-i \gamma a}). \label{eq:system} 
\end{eqnarray}
The system 
of equations \eqref{eq:system}
~has a nontrivial solution if the determinant of the $4\times 4$ matrix vanishes. By carrying out the calculation, 
one obtains
\begin{equation}\label{condition}
e^{4 i\gamma a}=\frac{f(\gamma)}{f^*(\gamma)}\equiv e^{i\zeta(\gamma)}\, ,
\end{equation}
with
\begin{equation}
f(\gamma)= \vert\varphi'(-a)\vert^2-\gamma^2\vert\varphi(-a)\vert^2
-2 i\gamma\Re[\varphi '(-a)\varphi^*(-a)],
\end{equation}
where the prime denotes the space coordinate ($x$) derivative and $\Re$ denotes the real part of the argument.

In the particular case in which $U(x)={\rm const}$, 
$f(\gamma)$ is real, 
~$\zeta(\gamma)=0$, 
~and~the condition for resonant transmission reads $\gamma a =n \pi/2$,  
 that is, 
$n\lambda/2 = 2 a$. This is the known textbook example.
In~the general case, the~condition \eqref{condition} gives 
\begin{equation}
\gamma a =\frac{n\pi}{2}+\frac{1}{4}\zeta(\gamma)\, .
\end{equation}
This is the scattering analog of the anti-squeezing situation described above, with~the region II 
 playing the role of the free evolution between time-dependent frequency evolutions of Section~\ref{sec2.2}.

\section{Unexciting Backgrounds in Quantum Field Theory and Reflectionless~Potentials} \label{sec5}

In homogeneous backgrounds, the~Fourier modes  with momentum $\vec k$  of a scalar field  satisfy a set of uncoupled  equations, which in 
many cases (including a scalar field in a Robertson Walker metric or a Yukawa-coupled scalar) take the form
\begin{equation}
\label{eq:mode_eqs}
\ddot \chi_k + \omega_k^2(t)\, \chi_k = 0,
\end{equation}
where $k=\vert \vec k\vert$.
For instance, for~massive and conformally coupled scalar fields in a Robertson–Walker metric, one has $\omega_k^2=k^2 +  m^2 a^2(t)$
where $a(t)$ is the scale factor 
for cosmological backgrounds~\cite{Birrell}. On~the other hand, in~scalar QED the frequency reads $\omega_k^2(t) = (k_\parallel - A_\parallel(t))^2 + k_\perp^2 + m^2\,$, where $A_\parallel(t)$ is a time-dependent 
component of the
vector potential in the case of an external electric field.  If~the quantum field is coupled to a classical field $\sigma(t)$ via  Yukawa coupling, the~squared frequency reads  $\omega_k^2(t) = k^2 + m^2 +\lambda \sigma^2(t)$.

In general, due to $k$-dependence, the~construction of STA via the Ermakov function \eqref{eq:fErmakov}
works only for individual modes, and~no global STA seems 
to exist unless  the frequency becomes time-independent (for example, for~massless conformally coupled fields in the cosmological case). However, when $\omega^2(t)\to \text{const}$ as $t\to \pm \infty$,  unexciting backgrounds do arise, and~correspond to reflectionless potentials in the analogous one-dimensional scattering problem~\cite{KayMoses}.  Hence, for~homogeneous backgrounds in QFT, 
one can construct 
evolutions that produce no particle excitations, i.e.,~unexciting backgrounds.  This has been unnoticed in 
the earlier studies 
on the subject~\cite{STAqft1,Vachaspati2022}. 

\subsection*{Construction of Unexciting~Evolutions}
In standard 
quantum mechanics, the~reflectionless potentials can be constructed by fixing the number and values of the energies of the bound states (assuming $V(x)<0$ and $V(x)\to 0$ for $x\to\pm\infty$). 
The Kay--Moses method, 
adapted for the QHO with time-dependent frequency, can be summarized as follows~\cite{KayMoses,AmJPhys2014}:
\begin{itemize}
\item 
 select $N$ positive numbers $\kappa_1>\kappa_2>$ \dots $>\kappa_N$; 
\item 
 compute the positive numbers $c_n$:
\begin{equation}
c_n^2 = 2 \kappa_n \prod_{m=1,m\neq n}^N\frac{\kappa_n+\kappa_m}{\vert\kappa_n-\kappa_m\vert}\text{;} 
\end{equation}
\item 
 compute the elements $C_{ij}$ of a matrix $\mathbb C$:
\begin{equation}
C_{ij}=\frac{c_ic_j}{\kappa_i+\kappa_j}e^{-(\kappa_i+\kappa_j)t}\text{;}
\end{equation}
\item 
 compute the frequency
\begin{equation}\label{freq_STA}
\omega_k^2(t)=\omega_0^2(k) + 2 \frac{d^2}{dt^2}\log\det(\mathbb{I} +\mathbb {C})\, .
 \end{equation}
\end{itemize}

In the scattering problem in quantum mechanics, 
 the $N$ numbers  $-\kappa_n$ are the eigenvalues of the bound states. For~the unexciting backgrounds, 
 the~$\kappa_n$ 
 are just $N$ parameters describing the freedom in the choice of the background.
The simplest and most 
known example 
is the potential obtained by choosing $\kappa_n=\kappa^2 n^2, \, \, n=1,2,\dots,N$ which
gives
\begin{equation}
\omega_k^2(t)=\omega_0^2(k) +  \frac{N(N+1)\kappa^2}{\cosh^2(\kappa t)}
\end{equation}
representing the Poschl--Teller~potential. 

This consideration
can been applied, for~example, to~particle creation in external time-dependent electric fields~\cite{electric1, electric2}, where particular profiles of $A_\parallel(t)$ can prevent particle production for selected values of $k_\parallel$, independently of $k_\perp$ and $m^2$. The~results of 
those studies
can be generalized by considering other profiles for the potential vector that produce a time-dependent frequency of the form Equation \eqref{freq_STA}.

Another interesting example is the case of a massive field in the $d$-dimensional de Sitter 
background 
\begin{equation}
\text{d}s^2 = \ell^2 (-\text{d}t^2 +\cosh^2t\,\,  \text{d}\Omega_{d-1})\, .
\end{equation}
Here, $d\Omega_{d-1}$ is the metric on the sphere $S^{d-1}$ and $\ell$ is related with the cosmological constant $\Lambda= (d-2)(d-1)/(2\ell^2)$. 
When expanding the modes of the field in generalized spherical harmonics, the~field equation
becomes 
\begin{equation}
\left[ \partial_t^2 +\frac{(2 l +n-3)}{2}\frac{(2 l +n-1)}{2}\frac{1}{\cosh^2t}+\left(m^2-\frac{(n-1)^2}{4}\right)\right]u_l(t)=0\, ,
\end{equation}
where   
$l$ is a non-negative integer. The~time-dependent part of the frequency coincides with the Poschl–Teller potential  if
$(2l+d-3)/2$ is a positive integer, and~this is the case in odd dimensions. This is another non-trivial example of an unexciting gravitational background, which has been pointed out in Refs.~\cite{deSitter1,deSitter2}.

These discussion highlights the  connection between STA in QFT and reflectionless potentials in 
standard
quantum mechanics, providing a unifying perspective on excitation-free evolutions in diverse quantum~systems.

\section{Conclusions} \label{sec6}
In this paper, 
we have analyzed the existence of STA and STA completions for the QHO with time-dependent frequency, as~well as for QFT in non-stationary backgrounds. A~central outcome of our analysis is based on 
the  
known analogy with one-dimensional 
Schr\"odinger 
equation~\cite{Marinov,Audretsch}: the implementation of STA in the QHO provides direct analogy with 
 transmission resonances in scattering problems, while the realization of STA in QFT with homogeneous backgrounds corresponds to the existence of reflectionless~potentials.

We have also exploited the connection between particle creation and squeezed states to show that STA completions naturally give rise to the notion of an anti-squeezing operator. This perspective provides a unified framework in which excitation suppression, particle creation, and~squeezing transformations are seen as deeply interconnected features of time-dependent quantum~systems.

Beyond the specific results, our analysis underscores the value of the time-dependent QHO as a bridge between different areas of physics. The~analogies identified here suggest possible extensions, from~the design of STA protocols in quantum technologies to the study of excitation-free backgrounds in cosmology and gravitational physics. In~this sense, STA and their completions may offer both conceptual insight and practical tools for controlling dynamical quantum processes in a broad variety of~contexts.

In particular, the~study of STA and their completions holds significant promise for applications in quantum thermodynamics, such as the design and implementation of quantum heat engines and quantum batteries. These systems inherently rely on the controlled manipulation of quantum states within finite-time protocols, where minimizing excitations and dissipation is crucial for optimal performance. STA techniques offer a systematic way to achieve such control, while STA completions ensure that the system returns to a ground or target state at the end of the process, even in nonadiabatic regimes. Notably, the~working medium in these devices can, 
actually,
be a quantum field, opening the door to novel thermodynamic regimes where field-theoretic effects play a central role. Furthermore, platforms like circuit quantum electrodynamics (cQED), which provide a highly tunable and coherent environment, are particularly well-suited for realizing these protocols experimentally. As~such, exploring STA in this context not only advances 
the
theoretical understanding but also paves the way for practical implementations of efficient and scalable quantum thermal~machines.


\vspace{6pt} 





\section*{Acknowledgments}This research
 was funded by Consejo Nacional de Investigaciones Cient\'{\i}ficas y
T\'ecnicas (CONICET), Universidad de Buenos Aires (UBA) and Universidad Nacional de Cuyo (UNCuyo).

\end{document}